\documentclass[prd,aps,nofootinbib,showpacs,showkeys,preprintnumbers]
{revtex4}
\usepackage{graphicx,epsf,amsmath,amsfonts,amssymb,amsbsy}
\usepackage{epsfig}
\newcommand{\ds}{\displaystyle}
\newcommand{\vev}[1]{\langle#1\rangle}
\newcommand{\mat}{\left ( \begin{array}}
\newcommand{\emat}{\end{array} \right )}
\newcommand{\vect}{\left ( \begin{array}{c}}
\newcommand{\evect}{\end{array} \right )}

\preprint{HU-EP-09/03}

\begin{document}

\title{ \bf Pion condensation in the Gross-Neveu model with nonzero
baryon and isospin chemical potentials}
\author{D.~Ebert $^1$ and K. G.~Klimenko $^2$}
\affiliation{$^1$ Institute of Physics, Humboldt-University Berlin,
10115 Berlin, Germany\\
$^2$ Institute
for High Energy Physics, 142281 Protvino, Moscow Region, Russia}

\begin{abstract}
The properties of the two-flavored Gross-Neveu model in the
(1+1)-dimensional spacetime are investigated in the presence of the
isospin $\mu_I$ as well as quark number $\mu$ chemical
potentials both at zero and nonzero temperatures. The consideration
is performed in the limit $N_c\to\infty$, i.e. in the case with
an infinite number of colored quarks. In the plane of parameters
$\mu_I,\mu$ a rather rich phase structure is found, which contains
phases with and without pion condensation. Moreover, we have found a
great variety of one-quark excitations of these phases, including
gapless and gapped quasiparticles.
\end{abstract}


\maketitle

\section{Introduction}

During the last decade great attention was paid to the
investigation of the QCD phase diagram in terms of baryonic
as well as isotopic (isospin) chemical potentials. First of
all, this interest is motivated by experiments on heavy-ion
collisions, where we have to deal with dense baryonic matter which
has an evident isospin asymmetry, i.e. different neutron and proton
contents of initial ions. Moreover, the dense hadronic/quark  matter
inside compact stars is also isotopically asymmetric. Generally
speaking, one of the important QCD applications is just to describe
the dense and hot baryonic matter. However, in the above mentioned
realistic situations the density is rather small, and weak coupling
QCD analysis is not applicable. So, different nonperturbative methods
or effective theories such as chiral effective Lagrangians and
especially  Nambu -- Jona-Lasinio (NJL) type models \cite{njl}
are usually employed for the consideration of the properties of dense
and hot baryonic matter under heavy-ion experimental and/or compact
star conditions, i.e. in the presence of such external conditions as
temperature and chemical potentials, magnetic field, finite size
effects etc (see, e.g., the papers
\cite{2,asakawa,alford,klim,warringa,incera,zhao,ebert,vshivtsev} and
references therein). In particular, the color superconductivity
\cite{alford,klim} as well as charged pion condensation
\cite{son,frank,ek,jin,andersen,abuki} phenomena of dense quark
matter were investigated in the framework of these QCD-like effective
models.

It is necessary to note that an effective description of QCD in terms
of NJL models, i.e. through an employment of
four-fermionic theories in (3+1)-dimensional spacetime, is usually
valid only at {\it rather low} energies and densities. Besides, at
present time there is the consensus that another class of theories,
the set of (1+1)-dimensional Gross-Neveu (GN) type models
\cite{gn,ft}, can also be used for a good qualitative consideration
of the QCD properties {\it without any restrictions} on the
energy/density values, which is in an encouranging contrast with NJL
models. Indeed, the GN type models are renormalizable, the
asymptotic freedom and spontaneous chiral symmetry breaking
are another properties inherent both for QCD and GN theories etc. In
addition, the $\mu-T$ phase diagram is qualitatively
the same in QCD and GN model \cite{wolff,kgk1,barducci,chodos,thies}
(here $\mu$ is the quark number chemical potential and $T$ is the
temperature). The GN type models are also very suitable for the
description of physics in such quasi one-dimensional condensed
matter systems as polyacetylene \cite{caldas}.
Due to the relative simplicity of GN models in the leading order of
the large $N_c$-expansion ($N_c$ is the number of colored quarks),
their use is very convenient for the application of
nonperturbative methods in quantum field theory \cite{okopinska}.
Moreover, it is necessary to note that just in the leading order of
the large $N_c$-expansion the well known no-go theorem by
Mermin-Wagner-Coleman \cite{coleman} apparently forbidding the
spontaneous breaking of continuous symmetries in the considered
(1+1)-dimensional models is not valid \cite{barducci,chodos,thies}.

Thus, such phenomena of dense QCD as color superconductivity, where
the color group is broken spontaneously, and charged pion
condensation, where the spontaneous breaking of the continuous
isospin symmetry takes place, might be simulated in terms of
simpler (1+1)-dimensional GN type models (see, e.g.,
\cite{chodos} and \cite{ekzt}, correspondingly).

In our previous paper \cite{ekzt} the phase diagram of the
(1+1)-dimensional GN model with two massless quark flavors was
investigated under the constraint that quark matter occupies a finite
space volume (see also the relevant papers \cite{kim}). In
particular, there we have studied in the large $N_c$-limit the
charged pion condensation phenomenon in cold quark matter with zero
baryonic density, i.e. at $\mu=0$, but nonzero
isotopic density, i.e. with nonzero isospin chemical potential
$\mu_I$. In contrast, in the present paper in the leading order of
the $1/N_c$-expansion we consider the phase portrait of the above
mentioned GN model in a more general case, where both isospin- and
quark number chemical potentials are nonzero, i.e. $\mu_I\ne 0$ and
$\mu\ne 0$. Moreover, we deal with nonzero temperature and with
spacetime of the usual topology, $R^1\times R^1$. We hope that our
investigations will shed some new light on the physics of dense and
hot isotopically asymmetric quark matter which can be observed in
heavy-ion collision experiments or might exist in compact stars,
where baryon density is obviously nonzero (i.e. $\mu\ne 0$). Our
consideration is based, for simplicity, on the approach with
homogeneous condensates (an extension to inhomogeneous
condensates was also considered in \cite{thies,dunne}).

The paper is organized as follows. In Section II the thermodynamic
potential of the two-flavored Gross-Neveu model is obtained in the
presence of quark number as well as isotopic chemical potentials.
In Section III and IV the phase structure of the model is
investigated at zero and nonzero temperatures, correspondingly. It
turns out that at zero temperature, $\mu_I\ne 0$ and rather small
values of $\mu$ the gapped pion condensed phase occurs. However,
at larger values of $\mu$ the dense quark matter phase with chirally
broken phase and gapless quasiparticles is realized.

\section{ The model and its thermodynamic potential}

We consider a (1+1)-dimensional model which describes dense quark
matter with two massless quark flavors ($u$ and $d$ quarks). Its
Lagrangian has the form
\begin{eqnarray}
&&  L=\bar q\Big [\gamma^\nu\mathrm{i}\partial_\nu
+\mu\gamma^0+\frac{\mu_I}2 \tau_3\gamma^0\Big ]q+ \frac {G}{N_c}\Big
[(\bar qq)^2+(\bar q\mathrm{i}\gamma^5\vec\tau q)^2 \Big ],
\label{1}
\end{eqnarray}
where the quark field $q(x)\equiv q_{i\alpha}(x)$ is a flavor
doublet ($i=1,2$ or $i=u,d$) and color $N_c$-plet
($\alpha=1,...,N_c$) as well as a two-component Dirac spinor (the
summation in (\ref{1}) over flavor, color, and spinor indices is
implied); $\tau_k$ ($k=1,2,3$) are Pauli matrices; the quark number
chemical potential $\mu$ in (\ref{1}) is responsible for the nonzero
baryonic density of quark matter, whereas the isospin chemical
potential $\mu_I$ is taken into account in order to study properties
of quark matter at nonzero isospin densities (in this case the
densities of $u$ and $d$ quarks are different). Evidently, the model
(\ref{1}) is a generalization of the (1+1)-dimensional Gross-Neveu
model \cite{gn} with a single massless quark color $N_c$-plet to the
case of two quark flavors and additional chemical potentials. As a
result, in the considered case we have a modified chiral symmetry
group. Indeed, at $\mu_I =0$ apart from the color SU($N_c$)
symmetry, the Lagrangian (\ref{1}) is invariant under
transformations from the chiral $SU_L(2)\times SU_R(2)$ group.
However, at $\mu_I \ne 0$ this symmetry is reduced to
$U_{I_3L}(1)\times U_{I_3R}(1)$, where $I_3=\tau_3/2$ is the third
component of the isospin operator (here and above the subscripts
$L,R$ mean that the corresponding group acts only on left, right
handed spinors, respectively). Evidently, this symmetry can also be
presented as $U_{I_3}(1)\times U_{AI_3}(1)$, where $U_{I_3}(1)$ is
the isospin subgroup and $U_{AI_3}(1)$ is the axial isospin
subgroup. Quarks are transformed under these subgroups as $q\to\exp
(\mathrm{i}\alpha\tau_3) q$ and $q\to\exp (\mathrm{i}
\alpha\gamma^5\tau_3) q$, respectively.

The linearized version of the Lagrangian (\ref{1}), which contains
composite bosonic fields $\sigma (x)$ and $\pi_a (x)$ $(a=1,2,3)$,
has the following form:
\begin{eqnarray}
\tilde L\ds &=&\bar q\Big [\gamma^\nu\mathrm{i}\partial_\nu
+\mu\gamma^0+ \frac{\mu_I}2\tau_3\gamma^0-\sigma
-\mathrm{i}\gamma^5\pi_a\tau_a\Big ]q
 -\frac{N_c}{4G}\Big [\sigma\sigma+\pi_a\pi_a\Big ].
\label{2}
\end{eqnarray}
From the Lagrangian (\ref{2}) one gets the following constraint
equations for the bosonic fields
\begin{eqnarray}
\sigma(x)=-2\frac G{N_c}(\bar qq);~~~\pi_a (x)=-2\frac G{N_c}(\bar q
\mathrm{i}\gamma^5\tau_a q). \label{200}
\end{eqnarray}
Obviously, the Lagrangian (\ref{2}) is equivalent to the Lagrangian
(\ref{1}) when using the constraint equations (\ref{200}).
Furthermore, it is clear from (\ref{200}) that the bosonic fields
transform under the isospin $U_{I_3}(1)$ and axial isospin
$U_{AI_3}(1)$ subgroups in the following manner:
\begin{eqnarray}
U_{I_3}(1):&&\sigma\to\sigma;~~\pi_3\to\pi_3;~~\pi_1\to\cos
(2\alpha)\pi_1+\sin (2\alpha)\pi_2;~~\pi_2\to\cos
(2\alpha)\pi_2-\sin (2\alpha)\pi_1,\nonumber\\
U_{AI_3}(1):&&\pi_1\to\pi_1;~~\pi_2\to\pi_2;~~\sigma\to\cos
(2\alpha)\sigma+\sin (2\alpha)\pi_3;~~\pi_3\to\cos
(2\alpha)\pi_3-\sin (2\alpha)\sigma. \label{201}
\end{eqnarray}
Due to the transformation rules (\ref{201}) such quantity as the
thermodynamic potential (TDP) depends effectively only on the two
combinations ($\sigma^2+\pi_3^2$) and ($\pi_1^2+\pi_2^2$) of the
bosonic fields, which are invariants with respect to the
$U_{I_3}(1)\times U_{AI_3}(1)$ group. In this case, without loss of
generality, one can put $\pi_2=\pi_3=0$, and study the TDP as a
function of only two variables, $M\equiv\sigma$ and
$\Delta\equiv\pi_1$. Throughout the paper we suppose for simplicity
that the condensates $M$ and $\Delta$ are homogeneous quantities
which do not depend on the space coordinate. In order to avoid the
no-go theorem \cite{coleman}, which forbids the spontaneous breaking
of continuous symmetries in the considered case of one space
direction, all our considerations are performed in the leading order
of the $1/N_c$-expansion, i.e. in the limit $N_c\to\infty$, where
the TDP of the model looks like (see, e.g., \cite{ekzt})
\begin{eqnarray}
\Omega_{\mu,\nu}(M,\Delta)
=\frac{M^2+\Delta^2}{4G}+\mathrm{i}\int\frac{d^2p}{(2\pi)^2}\ln
\Big\{\Big [(p_0+\mu)^2-(E^+_{\Delta})^2\Big ]\Big
[(p_0+\mu)^2-(E^-_{\Delta})^2\Big ]\Big\}, \label{9}
\end{eqnarray}
where $E_\Delta^\pm=\sqrt{(E^\pm)^2+\Delta^2}$, $E^\pm=E\pm\nu$,
$\nu=\mu_I/2$ and $E=\sqrt{p_1^2+M^2}$. It is clear that the TDP
$\Omega_{\mu,\nu}(M,\Delta)$ is symmetric under the transformations
$\mu\to -\mu$ and/or $\nu\to-\nu$. So it is sufficient to consider
only the region $\mu\geq 0$, $\nu\geq 0$. Taking into account this
constraint as well as integrating in (\ref{9}) over $p_0$, we obtain
for the TDP of the system at zero temperature the following
expression:
\begin{eqnarray}
\Omega_{\mu,\nu}(M,\Delta)&&\!\!\!\!\!\!
=\frac{M^2+\Delta^2}{4G}-\int_{-\infty}^{\infty}\frac{dp_1}{2\pi}\Big
\{E^+_{\Delta}+E^-_{\Delta}+(\mu-E^+_{\Delta})\theta
(\mu-E^+_{\Delta})+(\mu-E^-_{\Delta})\theta
(\mu-E^-_{\Delta})\Big\}, \label{12}
\end{eqnarray}
where $\theta (x)$ is the Heaviside step function. Since we are
going to study the phase diagram of the initial GN model, the system
of gap equations is needed:
\begin{eqnarray}
0=\frac{\partial\Omega_{\mu,\nu}(M,\Delta)}{\partial M}&\equiv&
\frac{M}{2G}-M\int_{-\infty}^{\infty}\frac{dp_1}{2\pi
E}\Big\{\frac{\theta(E_\Delta^+-\mu)E^+}{E_\Delta^+}+
\frac{\theta(E_\Delta^--\mu)E^-}{E_\Delta^-} \Big\},\nonumber\\
0=\frac{\partial\Omega_{\mu,\nu} (M,\Delta)}{\partial\Delta}&\equiv&
\frac{\Delta}{2G}-\Delta\int_{-\infty}^{\infty}\frac{dp_1}{2\pi}\Big
\{\frac{\theta(E_\Delta^+-\mu)}{E_\Delta^+}+
\frac{\theta(E_\Delta^--\mu)}{E_\Delta^-} \Big\}. \label{13}
\end{eqnarray}
Evidently, the coordinates $M$ and $\Delta$ of the global minimum
point of the TDP (\ref{12}) supply us with two order parameters
(gaps), which are proportional to the ground state expectation
values $\vev{\bar qq}$ and $\vev{\bar q\mathrm{i}\gamma^5\tau_1 q}$,
respectively. If only the gap $M$ is nonzero, then in the ground
state of the model the axial isospin symmetry $U_{AI_3}(1)$ (at
$\mu_I\ne 0$) is spontaneously broken down. However, if only the gap
$\Delta\ne 0$, then the ground state describes the phase with
charged pion condensation, where the isospin $U_{I_3}(1)$ symmetry
is spontaneously broken. Note that in this phase the space parity is
also spontaneously broken.

\section{Phase portrait at zero temperature}

First of all, let us consider the phase portrait of the initial GN
model (1) using as a starting point the TDP (\ref{12}). Since it is
an ultraviolet divergent quantity, one should renormalize it, using
a special dependence of the bare coupling constant $G\equiv
G(\Lambda)$ on the cutoff parameter $\Lambda$ ($\Lambda$ restricts
the integration region in the divergent integrals, $|p_1|<\Lambda$).
In our previous paper \cite{ekzt} the following prescription for the
bare coupling constant $G(\Lambda)$ was used,
\begin{eqnarray}
\frac{1}{2G(\Lambda)}=\frac{1}{\pi}\int_{-\Lambda}^\Lambda
dp_1\frac{1}{\sqrt{M_0^2+p_1^2}}=\frac{2}{\pi}\ln\left
(\frac{\Lambda+\sqrt{M_0^2+\Lambda^2}}{M_0}\right ), \label{16}
\end{eqnarray}
where $M_0$ is the dynamically generated quark mass in the vacuum,
i.e. at $\mu=0$ and $\mu_I=0$ (see below). Then, introducing the
quantity $\Omega_{\mu,\nu}(M,\Delta;\Lambda)$,
\begin{eqnarray}
\Omega_{\mu,\nu}(M,\Delta;\Lambda)&&\!\!\!\!\!\!
=\frac{M^2+\Delta^2}{4G(\Lambda)}-\int_{-\Lambda}^{\Lambda}
\frac{dp_1}{2\pi}\Big\{E^+_{\Delta}+E^-_{\Delta}+(\mu-E^+_{\Delta})
\theta(\mu-E^+_{\Delta})+(\mu-E^-_{\Delta})\theta
(\mu-E^-_{\Delta})\Big\}+\frac{\Lambda^2}{\pi}, \label{120}
\end{eqnarray}
it is possible to obtain the renormalized (finite) expression for
the TDP:
\begin{eqnarray}
\Omega_{\mu,\nu}(M,\Delta)=\lim_{\Lambda\to\infty}\Omega_{\mu,\nu
}(M,\Delta;\Lambda). \label{1201}
\end{eqnarray}
(The renormalized expression for the gap equations is obtained in
the limit $\Lambda\to\infty$, if the replacements $G\to G(\Lambda)$
and $|p_1|<\Lambda$ are done in (\ref{13}), or by a direct
differentiation of the expression (\ref{1201}).) In particular, at
$\mu=0$ and $\mu_I=0$ we have from (\ref{1201}):
\begin{eqnarray}
\Omega_{\mu,\nu}(M,\Delta){\Big |}_{\mu=0,\nu=0}\equiv
V_0(\sqrt{M^2+\Delta^2})=\frac{M^2+\Delta^2}{2\pi}\left [\ln\left
(\frac{M^2+\Delta^2}{M_0^2}\right )-1\right ]. \label{17}
\end{eqnarray}
Since for a strongly interacting system the space parity is expected
to be a conserved quantity in the vacuum, we put $\Delta$ equal to
zero in (\ref{17}). As a result, the global minimum of the TDP
(\ref{17}) (usually, this quantity is called effective potential in
the vacuum) lies in the point $M=M_0$, which means that in the
vacuum the dynamically generated quark mass is just the parameter
$M_0$ introduced in (\ref{16}). However, in the general case, i.e.
at nonzero values of the chemical potentials, this quantity depends
certainly on $\mu,\mu_I$ and obeys the gap equations (\ref{13}).

Numerical investigations show that local minima of the TDP
(\ref{1201}) can occur only on the $M$- or $\Delta$-axes (other
solutions of the gap system (\ref{13}) correspond to saddle points
of the TDP (\ref{1201})). So it is enough to restrict the
consideration of the TDP (\ref{1201}) to the regions $M=0,\Delta\ne
0$ ($\Delta$-axis) or $M\ne 0,\Delta=0$ ($M$-axis). Moreover, since
the TDP is an even function with respect to the transformations
$M\to -M$ or $\Delta\to -\Delta$, we will suppose further that
$M,\Delta\geq 0$. As a result we have
\begin{eqnarray}
\Omega_{\mu,\nu}(M=0,\Delta)\!\!&=&\!\!
V_0(\Delta)-\frac{\nu^2}\pi+\frac{\theta(\mu-\Delta)}{\pi} \left
[\Delta^2\ln\left (\frac{\mu+\sqrt{\mu^2-\Delta^2}}{\Delta}\right
)-\mu\sqrt{\mu^2-\Delta^2}\right ],
\label{18}\\
\Omega_{\mu,\nu}(M,\Delta=0)\!\!&=&\!\!
V_0(M)+\frac{\theta(\mu+\nu-M)}{2\pi} \left [M^2\ln\left
(\frac{\mu+\nu+\sqrt{(\mu+\nu)^2-M^2}}{M}\right
)-(\mu+\nu)\sqrt{(\mu+\nu)^2-M^2}\right ]\nonumber\\
&&~~~~~~~~~~~~~~~~~~~~~+\frac{\theta(|\mu-\nu|-M)}{2\pi} \left
[M^2\ln\left (\frac{|\mu-\nu|+\sqrt{(\mu-\nu)^2-M^2}}{M}\right
)\right.\nonumber\\
&&\left.~~~~~~~~~~~~~~~~~~~~-|\mu-\nu|\sqrt{(\mu-\nu)^2-M^2}\right ]
\Big (1-\theta(M-\nu)\theta(\nu-\mu)\Big ), \label{19}
\end{eqnarray}
where the function $V_0(x)$ is presented in (\ref{17}). Compairing
the global minima of the TDPs (\ref{18})-(\ref{19}), it is possible
to obtain the global minimum point of the genuine TDP (\ref{1201})
of the model. Clearly, its form (and, as a result, the phase of the
model) depends on the values of $\mu$ and $\nu$. So, the total
information about the behaviour of the global minimum point vs
$\mu$, $\nu$ can be presented by the phase portrait given in Fig. 1.
(Note, it is valid only for $\nu > 0$ values. In the $\nu=0$ case
the phase structure is described below). There, in the corresponding
$(\tilde\nu,\tilde\mu)$-plane (excluding the $\tilde\mu$-axis),
where $\tilde\mu=\frac{\mu}{M_0}$, $\tilde\nu=\frac{\nu}{M_0}
=\frac{\mu_I}{2M_0}$, you can see three respective phase regions
denoted by the numbers 1, 2, 3. For the values of
$(\tilde\nu,\tilde\mu)$ from the region 1, 2, and 3 the global
minimum point of the TDP (\ref{1201}) has the form $(M=0,\Delta=0)$,
$(M\ne 0,\Delta=0)$, and $(M=0,\Delta\ne 0)$, correspondingly. As a
result, in the region 1 the chirally $U_{I_3}(1)\times
U_{AI_3}(1)$-symmetric phase with massless quarks is arranged.  In
the region 2, where the order parameter $M$ is nonzero, this
symmetry is spontaneously broken down to the isospin $U_{I_3}(1)$
subgroup (in this region the order parameter $M$ is a smooth
function vs $\mu$ and $\nu$). We call this phase the normal quark
matter phase, since here quarks dynamically acquire a mass which is
equal to the order parameter $M$, and space parity is not broken.
Finally, the region 3 corresponds to the charged pion condensed
phase, because of the nonzero order parameter $\Delta$. Furthermore,
for all points of this region $\Delta\equiv M_0$. Note that on the
phase boundaries phase transitions of the first order occur.
\begin{figure}
 \includegraphics[width=0.45\textwidth]{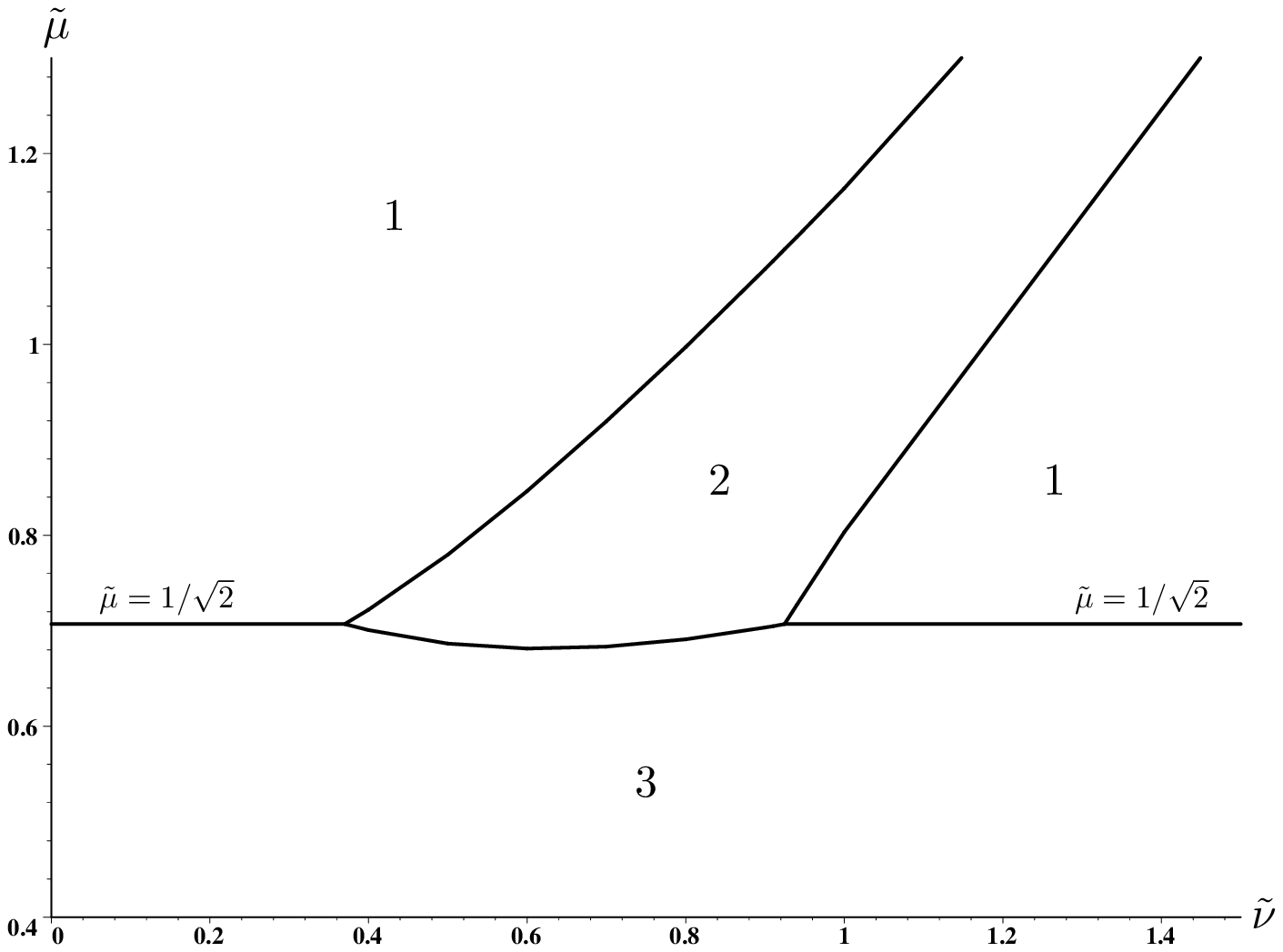}
 \hfill
 \includegraphics[width=0.45\textwidth]{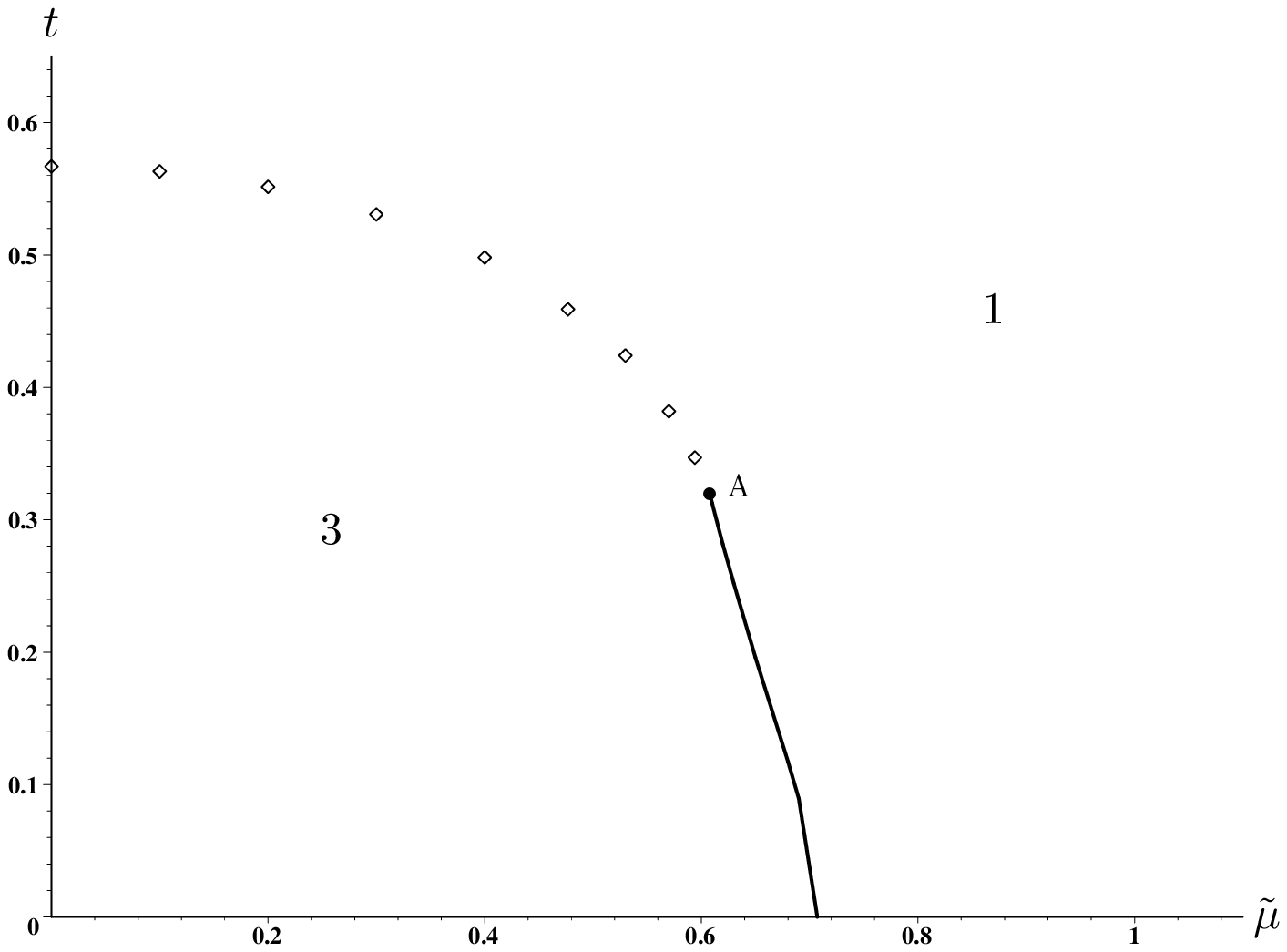}\\
\parbox[t]{0.45\textwidth}{
 \caption{The $(\tilde\mu,\tilde\nu)$ phase portrait of the model at
 $T=0$ and $\tilde\nu>0$.
  Here  $\tilde\mu=\frac{\mu}{M_0}$,  $\tilde\nu=\frac{\mu_I}{2M_0}$,
 and  $M_0$ is the quark mass in the vacuum. Number 1 denotes the
 symmetric phase with massless quarks, number 2 -- the normal quark
 matter phase with massive quarks,
 and 3 denotes the pion condensed phase.} \label{fig:1} }
\hfill
\parbox[t]{0.45\textwidth}{
\caption{The $(t,\tilde\mu)$  phase portrait of the model at
$\mu_I=0.2M_0$ ($t=\frac{T}{M_0},\tilde\mu=\frac{\mu}{M_0}$). The
solid and dotted lines are the critical curves of the 1st and 2nd
order phase transitions, correspondingly. They are divided by the
tricritical point A. Other notions are the same as in Fig. 1.}
\label{fig:2} }
\end{figure}

It is worth to remark that in early papers (see e.g.
\cite{wolff,kgk1}), the phase structure of the GN model was studied
in the exceptional case of $\nu=0,\mu\ne 0$ with homogeneous
condensates. As it was shown there, at $\mu>M_0/\sqrt{2}$ the
symmetric phase 1 is observed. However, at $\mu<M_0/\sqrt{2}$ the
normal quark matter phase is realized with a constant quark mass
equal to $M_0$. \footnote{If an inhomogeneous ansatz for condensates
is taken into account, then in some GN phase diagrams an additional
crystalline phase might appear at $\nu=0$ and intermediate values of
$\mu$ \cite{thies,dunne}. The existence of the crystalline phase in
the general case with $\nu\ne 0$ is not investigated up to now.} Let
us denote this particular normal phase as phase 4, which is not
depicted in Fig. 1. \footnote{We would like to stress once more that
in this figure the phase 3 does not occupy the part of the
$\tilde\mu$-axis $(\tilde\nu=0$, $0<\tilde\mu<1/\sqrt{2})$, where
actually the phase 4 is arranged.} There are two essential
differences between phases 2 and 4. The first one manifests itself
in their dynamical properties such as quasiparticle excitations.
Recall that in the most general case the energy spectrum of the
$u$-, $d$-, $\bar u$-, $\bar d$-quasiparticles (quarks) has the
following form (see, e.g., \cite{ek}):
\begin{eqnarray}
E_u(p_1)=E_\Delta^--\mu,~~E_d(p_1)=E_\Delta^+-\mu,~~E_{\bar u}(p_1)
=E_\Delta^++\mu,~~E_{\bar d}(p_1)=E_\Delta^-+\mu. \label{20}
\end{eqnarray}
It is easily seen from (\ref{20}) that in the phase 4, where
$\nu=0$, $\Delta=0$, $M=M_0$ and $\mu<M_0/\sqrt{2}$, there is a gap
in the quasiparticle energy spectrum, i.e. in the phase 4 all
quasiparticles are gapped excitations. In contrast, in the phase 2
($\nu\ne 0$) the situation is quite different and here in the energy
spectrum of the $u$-quasiparticles the gap is absent. It means that
for each point of the region 2 in Fig. 1 there exists a space
momentum $p_1^\star$ such that $E_u(p_1^\star)=0$. (For example, the
point $(\tilde\nu=0.7,\tilde\mu\approx 0.92)$ is in the phase 2
(near its boundary), where $\Delta=0$ and $M\approx 0.31 M_0$. In
this case $p_1^\star\approx 1.59 M_0$ etc.) So, in analogy with
gapped and gapless color superconductivity \cite{alford}, the phase
2 may be called gapless normal quark matter phase.

The second difference between phases 2 and 4 is a thermodynamic one
and related with the density of quarks in both phases. It is clear
that in the phase 4, i.e. at $\nu=0$, $M=M_0$ and
$\mu<M_0/\sqrt{2}$, the TDP from (\ref{19}) does not depend on the
quark chemical potential $\mu$. As a result, we see that in this
phase the quark density $n_q\equiv -\partial\Omega/\partial\mu$ is
equal to zero. However, in the phase 2 the density of quarks is not
zero. So the consideration of dense quark matter in the framework of
the GN model is more adequate in terms of both nonzero isotopic and
quark number chemical potentials, $\nu\ne 0$ and $\mu\ne 0$, than in
the simpler case with $\nu=0$, $\mu\ne 0$.

It is seen from Fig. 1 that the symmetric phase 1, where $M=0$ and
$\Delta =0$, is a disconnected manifold consisting of two different
parts. (Note that in the ground state of this phase the quark number
density is not zero.) The first part is arranged above the phase 2
and corresponds to $\mu>\nu$, whereas in the second one, which  is
below the phase 2, we have $\mu<\nu$. It turns out that the
dynamical properties of these two regions are different. Indeed, for
the part of the phase 1 which is above the phase 2  both $u$- and
$d$-quasiparticles are gapless. However, in the rest of the phase 1,
which corresponds to $\mu<\nu$, only $u$-quasiparticles are gapless,
as it is clear from the corresponding dispersion relations of these
quasiparticles, $E_u(p_1)=|p_1-\nu|-\mu$ and $E_d(p_1)=p_1+\nu-\mu$,
in the phase 1. Hence, such dynamical properties of dense quark
matter as transport phenomena (e.g., conductivities etc) can occur
in a qualitatively different way in these subphases of the symmetric
phase 1.

Finally, note that all quasiparticle excitations of the pion
condensed phase 3 are gapped ones. Indeed, it is clear from
(\ref{20}) that at $\Delta =M_0$, $M=0$ and $\mu\le M_0/\sqrt{2}$
the quantities (\ref{20}) do not vanish. So there is a gap in the
energy spectrum of quasiparticles in the pion condensed phase 3.

\section{Phase portrait at nonzero temperatures}

Now let us stydy the influence of the temperature $T$ on the phase
structure of the considered GN model with two chemical potentials
$\mu$ and $\nu\equiv\mu_I/2$. To get the corresponding thermodynamic
potential $\Omega_{\mu,\nu,\scriptscriptstyle{T}}(M,\Delta)$, one
can simply start from the expression for the TDP at zero temperature
(\ref{9}) and perform the following standard replacements:
\begin{eqnarray}
\int_{-\infty}^{\infty}\frac{dp_0}{2\pi}\big (\cdots\big )\to
iT\sum_{n=-\infty}^{\infty}\big (\cdots\big ),~~~~p_{0}\to
p_{0n}\equiv i\omega_n \equiv i\pi T(2n+1),~~~n=0,\pm 1, \pm 2,...,
\label{190}
\end{eqnarray}
i.e. the $p_0$-integration should be replaced in favour of the
summation over an infinite set of Matsubara frequencies $\omega_n$.
Summing in the obtained expression over Matsubara frequencies (the
corresponding technique is presented, e.g., in \cite{jacobs}), one
can find for the TDP:
\begin{eqnarray}
\Omega_{\mu,\mu_I,\scriptscriptstyle{T}}(M,\Delta)\!
&=&\frac{M^2+\Delta^2}{4G}-\int_{-\infty}^{\infty}\frac{dp_1}{2\pi}
\Big\{E^+_{\Delta}+E^-_{\Delta}+T\ln\big [1+e^{-\beta
(E^+_{\Delta}-\mu)}\big ]+T\ln\big [1+e^{-\beta
(E^+_{\Delta}+\mu)}\big ]\nonumber\\ &+&T\ln\big [ 1+e^{-\beta
(E^-_{\Delta}-\mu)}\big ]+T\ln\big [1+e^{-\beta
(E^-_{\Delta}+\mu)}\big ]\Big\}, \label{1202}
\end{eqnarray}
where $\beta=1/T$. As in the case with zero temperature, the
renormalized expression for
$\Omega_{\mu,\nu,\scriptscriptstyle{T}}(M,\Delta)$ can be obtained
by the replacement $G\to G(\Lambda)$ (see formula (\ref{16})) which,
along with the cutting of the integration region in (\ref{1202}),
$|p_1|<\Lambda$, leads in the limit $\Lambda\to\infty$ to the finite
expression for the TDP. Numerical investigations show that all
possible local minima of the obtained TDP lie on the $M$- and
$\Delta$-axis. So it is sufficient to deal with corresponding
restrictions of the TDP on these axes, i.e. with the following
functions,
\begin{eqnarray}
\Omega_{\mu,\mu_I,\scriptscriptstyle{T}}(M=0,\Delta)\!
&=&V_0(\Delta)-\frac{\nu^2}{\pi}
-\frac{2T}{\pi}\int_{0}^{\infty}dp_1\ln \Big\{\left [1+e^{-\beta
({\cal E}-\mu)}\right ]\left [1+e^{-\beta
({\cal E}+\mu)}\right ]\Big\},\label{1203}\\
\Omega_{\mu,\mu_I,\scriptscriptstyle{T}}(M,\Delta=0)\!
&=&V_0(M)-\frac{T}{\pi}\int_{0}^{\infty}dp_1\ln \Big\{\left
[1+e^{-\beta (E+\nu-\mu)}\right ]\left [1+e^{-\beta
(E+\nu+\mu)}\right ]\Big\}\nonumber\\
&&~~~~~~~~~-\frac{T}{\pi}\int_{0}^{\infty}dp_1\ln \Big\{\left
[1+e^{-\beta (E-\nu-\mu)}\right ]\left [1+e^{-\beta
(E-\nu+\mu)}\right ]\Big\},\label{1204}
\end{eqnarray}
where the effective potential $V_0(x)$ is given in
(\ref{17}), $E=\sqrt{p_1^2+M^2}$, and ${\cal
E}=\sqrt{p_1^2+\Delta^2}$. Comparing the global minima of the
functions (\ref{1203}) and (\ref{1204}), it is possible to establish
the global minimum point of the renormalized TDP (\ref{1202}). Then,
the dependence of the global minimum point vs $T,\mu,\nu$ defines
the phase structure of the model.

Using this prescription in our numerical investigations of the TDPs
(\ref{1203})-(\ref{1204}), we have found the three $(\mu,T)$-phase
portraits of the initial GN model, depicted in Figs. 2, 3, 4, for
three qualitatively different values of the isospin chemical
potentials, $\mu_I=0.2M_0$, $\mu_I=1.2M_0$, and $\mu_I=2M_0$,
respectively. There, the solid and dotted lines correspond to curves
of first- and second order phase transitions. Moreover, there are
several tricritical points, A, B, C, in these phase diagrams. ( A
point of the phase diagram is called a tricritical one, if an
arbitrary small vicinity of it contains both first- and second order
phase transition points.)
\begin{figure}
 \includegraphics[width=0.45\textwidth]{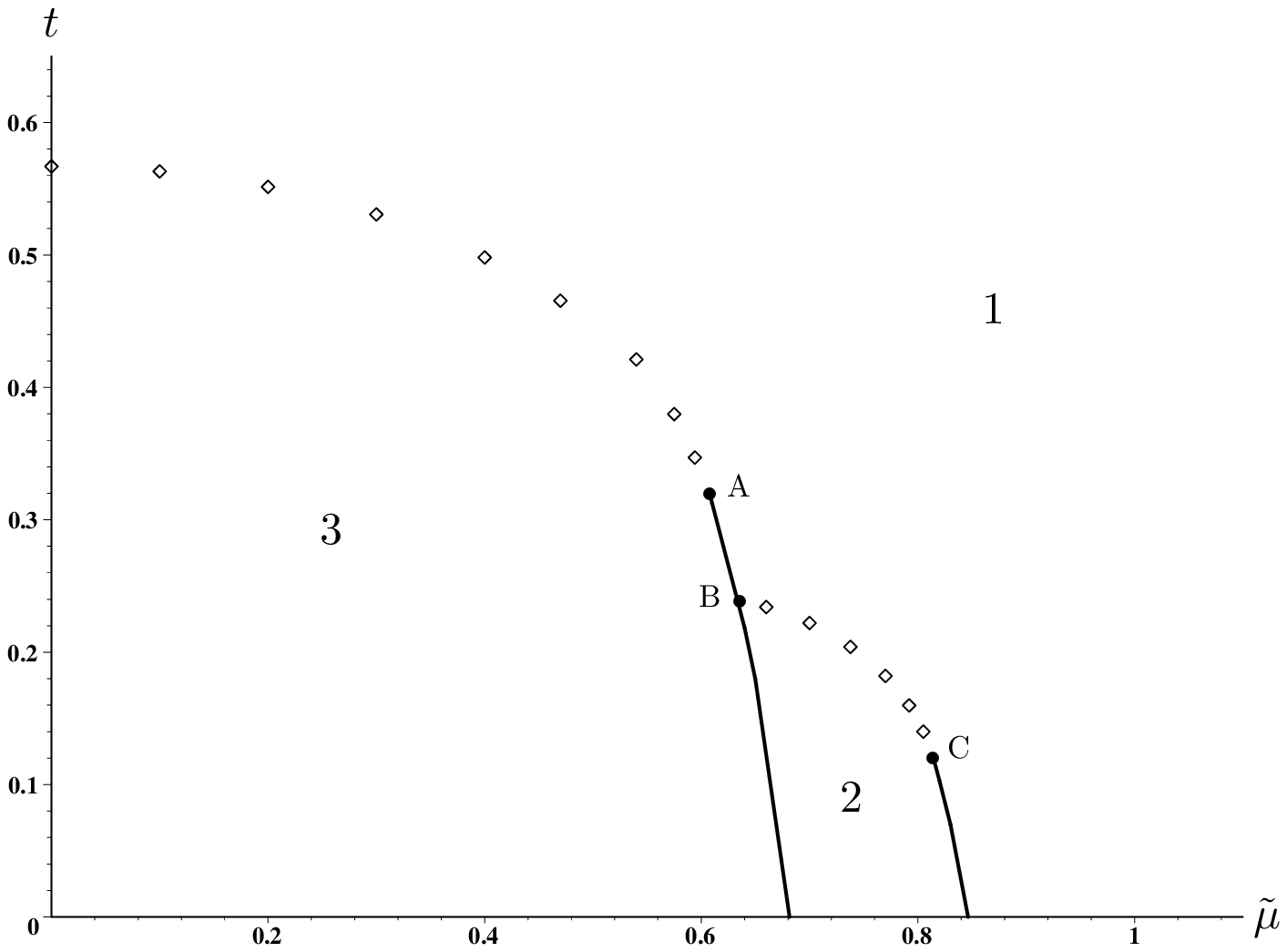}
 \hfill
 \includegraphics[width=0.45\textwidth]{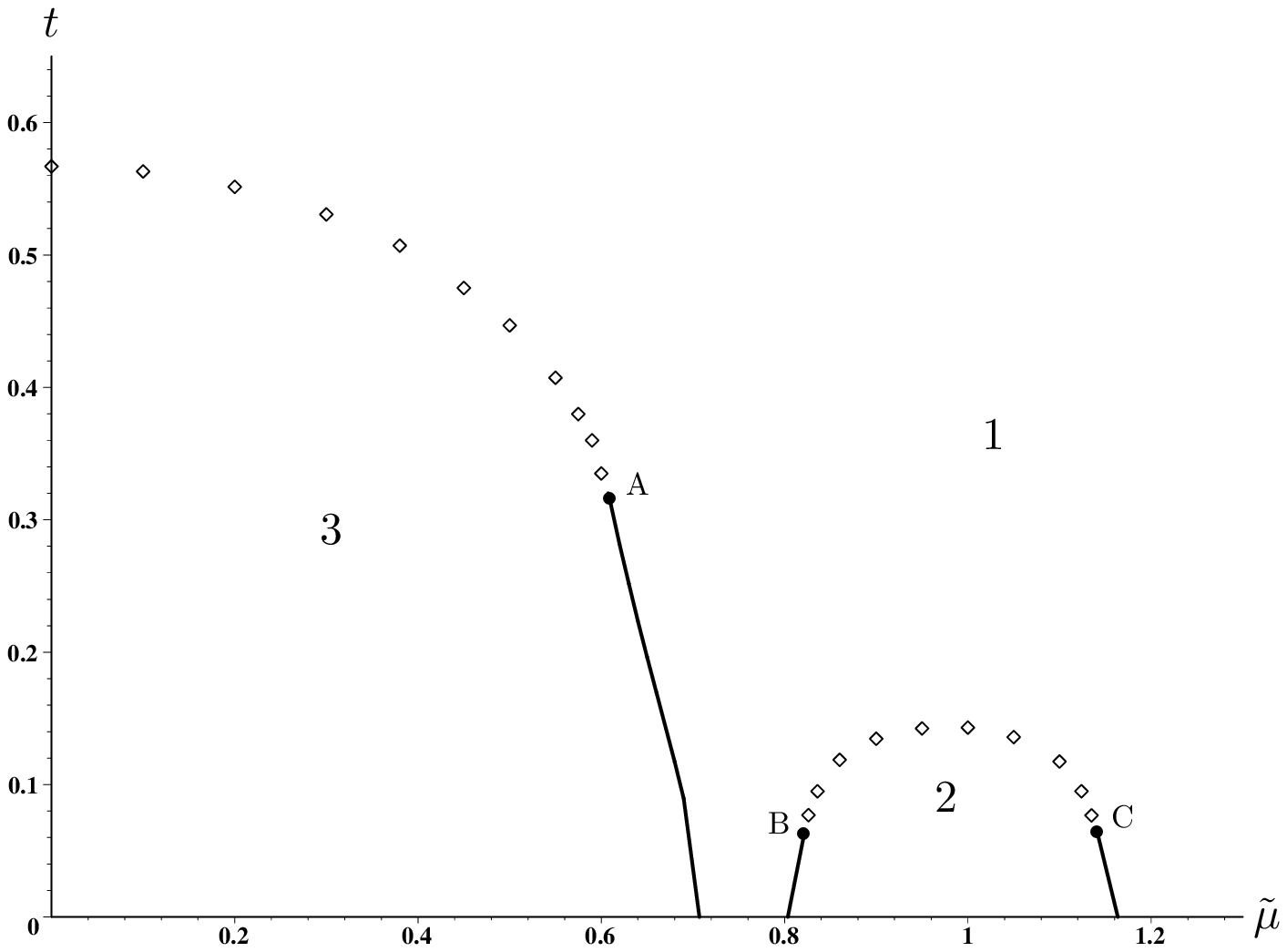}\\
\parbox[t]{0.45\textwidth}{
 \caption{The $(t,\tilde\mu)$  phase portrait
of the model at $\mu_I=1.2M_0$ ($t=\frac{T}{M_0}$,
$\tilde\mu=\frac{\mu}{M_0}$). The points A, B, C are tricritical
points. Other notions are the same as in Figs. 1, 2.} \label{fig:3}
} \hfill
\parbox[t]{0.45\textwidth}{
\caption{The $(t,\tilde\mu)$  phase portrait of the model at
$\mu_I=2M_0$ ($t=\frac{T}{M_0}$, $\tilde\mu=\frac{\mu}{M_0}$). Other
notions are the same as in the previous figures.} \label{fig:4} }
\end{figure}

\section{Summary and conclusions}

Recent investigations of the phase diagram of isotopically
asymmetric dense quark matter in terms of NJL models show that their
pion condensation content is not yet fully understood. Indeed, the
number of the charged pion condensation phases of the phase diagram
depends strictly on the parametrization set of the NJL model. It
means that for different values of the coupling constant, cutoff
parameter etc just the same NJL model predicts different numbers of
pion condensation phases of quark matter both with or without
electric neutrality constraint (see, e.g., \cite{ek,andersen}). So
to obtain a more objective information about the pion condensation
phenomenon of dense quark matter it is important to invoke
alternative approaches. One of them, which qualitatively quite
successfully imitates the QCD properties (see also the
Introduction), is based on the consideration of this phenomenon in
terms of the leading order of the large $N_c$-technique in the
framework of (1+1)-dimensional GN models.

In the present paper we have studied the phase structure of the GN
model (1) in terms of temperature, quark number ($\mu$)- as well as
isospin ($\mu_I$) chemical potentials in the limit $N_c\to\infty$.
Firstly, we have found that at $T=0$ the charged pion condensation
phase of the GN model is realized inside the chemical potential
region $\mu_I>0$, and $\mu$ is not greater than $M_0/\sqrt{2}$ (see
Fig. 1). The corresponding quasiparticle excitations have a gap in
the energy spectrum. In contrast, in the NJL phase diagram the pion
condensation phases occupy a compact region and for some
parameterization schemes the gapless pion condensation might occur
\cite{frank,ek,jin,andersen}.

Secondly, at $\mu_I>0$ and rather large values of the quark number
chemical potential $\mu$  we have found the normal quark matter
phase 2 (see Fig. 1) in which chiral symmetry is broken
spontaneously. Moreover, some of its one-quark excitations are
gapless quasiparticles, and the quark number density in the ground
state is not zero. (In comparision, in the early investigations of
the GN model at $\mu_I=0$ and $\mu\ne 0$ the chirally broken phase
was also found, but with zero quark number density and gapped
quasiparticles \cite{wolff}.) The region corresponding to this phase
in the $(\mu_I,\mu)$-plane is not restricted. In contrast, in the
chirally symmetric NJL model, i.e. with zero current quark mass, the
$(\mu_I,\mu)$-phase diagram does not contain the normal quark matter
phase with such properties (i.e. with nonzero quark number density,
gapless quasiparticles, and broken chiral invariance), at least at
$\mu_I>0$ \cite{ek}.

Thirdly, we have found a rather rich quasiparticle excitation
spectrum for the chirally symmetric phase 1. In this phase, if
$\mu>\nu$, all quasiparticles are gapless. However, if $\mu<\nu$,
then the gap is absent in the spectrum of only $u$-quasiparticles.

We hope that our investigation of the GN phase diagram will shed
some new light on the phase structure of QCD at nonzero baryonic and
isotopic densities.
Thus, even in the most simple approach to the GN phase diagram we
have found a variety of phases with rather rich dynamical contents.
Obviously, a more realistic imitation of the QCD phase diagram
requires to include also a nonzero bare quark mass, i.e. to study
massive GN models, as well as to take into account the possibility
of inhomogeneous condensates.

\end{document}